\documentclass[twocolumn]{aa}
\usepackage[switch]{lineno}
\usepackage{graphicx}
\usepackage{color}
\usepackage{txfonts}
\newcommand{\ii}{{\rm i}}
\newcommand{\dd}{{\rm d}}
\newcommand{\J}{\mathbf{J}}
\begin{document}

\title{The interaction of multiple stellar winds in stellar clusters: potential flow}
\titlerunning{Interaction of multiple stellar winds}

\author{
  K.~Scherer  \inst{1,2} \and
  A.~Noack    \inst{1} \and
  J.~Kleimann \inst{1} \and
  H.~Fichtner \inst{1,2} \and
  K.~Weis     \inst{3}
}

\institute{Institut f\"{u}r Theoretische Physik IV, Ruhr-Universit\"at
  Bochum, 44780 Bochum, Germany. \email{kls@tp4.rub.de}
  \and Research Department ``Plasmas with Complex Interactions'',
  Ruhr-Universit\"at Bochum, 44780 Bochum, Germany \and
  Astronomisches Institut, Ruhr-Universit\"at Bochum, 44780 Bochum, Germany
}

\date{Received ?, ?; accepted ?, ?}

% \abstract{}{}{}{}{} 
% 5 {} token are mandatory

% context heading (optional) {} leave it empty if necessary  
\abstract
    {While several studies have investigated large-scale cluster winds
      resulting from an intra-cluster interaction of multiple stellar winds,
       as yet they have not provided details of the bordering flows inside a
      given cluster.
    } % aims heading (mandatory)
    {The present work explores the principal structure of the combined
      flow resulting from the interaction of multiple stellar winds inside
      stellar clusters.
    } % methods heading (mandatory)
    {The theory of complex potentials is applied to analytically investigate
      stagnation points, boundaries between individual outflows, and the
      hydrodynamic structure of the asymptotic large-scale cluster wind.
      In a second part, these planar considerations are extended to fully
      three-dimensional, asymmetric configurations of wind-driving stars.
    } % results heading (mandatory)
    {We find (i) that one can distinguish regions in the large-scale
      cluster wind that are determined by the individual stellar winds, (ii)
      that there are comparatively narrow outflow channels, and (iii) that
      the large-scale cluster wind asymptotically approaches spherical symmetry
      at large distances. 
    } % conclusions heading (optional), leave it empty if necessary 
    {The combined flow inside a stellar cluster resulting from the interaction
      of multiple stellar winds is highly structured.
    }
    
%    \keywords{interacting stellar winds -- stellar clusters --
%      potential theory -- hydrodynamics}

      
      \maketitle
    
%    \fbox{Vorschlag \red{neuer Text in rot}}, \\
%    \fbox{Vorschlag \blue{wegfallender Text in blau}}.
    
% -----------------------------------------------------------------------------
\section{Introduction and motivation}
%------------------------------------------------------------------------------
%
According to a generally accepted paradigm, the structures known as superbubbles
\citep{Chu-2008, McClure-Griffith-2012, Ambrocio-Cruz-etal-2016} are the result
of the interaction of multiple stellar winds within stellar clusters.
Individual  examples are the central cluster at the Galactic center
\citep{Ozernoy-etal-1997}, the Arches cluster
\citep{Canto-etal-2000, Raga-etal-2001}, the Carina complex \citep{Gull-2011},
M~17 \citep{Reyes-Iturbide-etal-2009, Mernier-Rauw-2013}, 
and N~70 \citep{Rodriguez-Gonzalez-etal-2011}.

The interest in superbubbles is triggered by the desire to understand
their X-ray luminosities  \citep[e.g.,][]{Raga-etal-2001}, by their potential
contributions to the Galactic cosmic-ray flux 
\citep[e.g.,][]{Binns-etal-2007,Murphy-etal-2016} via acceleration at 
multiple shocks \citep{Bykov-2001, Ferrand-Marcowith-2010}, and because they are  sources of gamma rays \citep{Cesarsky-Montmerle-1983,
  Domingo-Santamaria-Torres-2006}.

In many cases the resulting large-scale wind outside a stellar cluster is 
assumed to be a spherically symmetric outflow, similar to that described 
by the original model of interstellar bubbles by \citet{Castor-etal-1975} and 
\citet{Weaver-etal-1977}. This assumption is  invalid inside a given
cluster where the winds of several (probably dominating) O and B stars
interact with each other. The mutual interaction of stellar winds has
been studied over many years in great detail for the case of binary stars 
\citep[e.g.,][and references therein]{Kallrath-1991, Pittard-2011, Reitberger-etal-2014, Reitberger_EA:2017}, including  
their additional interaction with the interstellar medium 
\citep{Wareing-etal-2007, Banda-Barragan-2016, Banda-Barragan-2018}.
Additionally,  \citet{Parkin_EA:2014} and \citet{Reitberger-etal-2014}, among others,  have studied the effects of orbital motion in binary systems. These studies are neglected here, however,  because we consider widely separated ($\sim 1$~pc) stars for which the orbital timescale is much longer ($\sim 10^7$~yrs) than the time taken for the interaction region of the winds to come into equilibrium ($\sim 10^5$~yrs).

Compared to the existing body of research on binary winds, the investigations of the simultaneous interaction of multiple stars are still in their infancy. 
The first quantitative model of a large-scale cluster wind resulting from a
more realistic internal flow structure was presented by
\citet{Canto-etal-2000}. For numerical purposes, these authors had to assume a
symmetric distribution of wind-driving stars inside the cluster. This model
 was subsequently generalized to a statistically homogeneous
\citep{Raga-etal-2001} and to a nonuniform 
\citep{Rodriguez-Gonzalez-etal-2007} stellar distribution.
These hydrodynamical (HD) simulations were later applied to explain the X-ray
emission of various clusters (as mentioned above) and further generalized to 
 include stellar winds and  a supernova event as well
\citep{Rodriguez-Gonzalez-etal-2011}. 

All of the models employing a nontrivial internal structure of
stellar clusters are of numerical nature and do not provide details of
the colliding plasma flows inside a given cluster. The present paper
intends to fill these modeling gaps. First, we present a simplified
two-dimensional (2D) analytical model of the interaction of multiple
stellar winds and the resulting large-scale cluster wind. Second, we
analyze the principal structure of such multisource wind flow
geometries. Third, we discuss the corresponding generalization to
explore asymmetric 3D configurations, sometimes resorting  to numerical methods.
In doing so, we restrict ourselves to stellar clusters where the mutual distances between the dominating wind-driving stars
  are large enough to justify neglect of any orbital motion.  Since
  the  termination shock surfaces of the stars  are disjoint, the entire wind
  interaction region is subsonic and thus, to a good
  approximation, can be taken as incompressible.  As shown by several
  authors \citep{Arthur-2007,Mackey-etal-2015,Scherer-etal-2016}, at
  least for isolated astrospheres, the density inside an astropause is
  so low that it is not affected by cooling. The region between
  astropause and bow shock is affected, but this is not relevant to
  our study because there are no bow shocks, but rather flows of the
  intracluster medium around the astropauses.  For these, we assume
  that this medium is in thermal equilibrium, i.e.,\ that cooling is not
  effective, and  thus it can be neglected as well.
% -----------------------------------------------------------------------------
\section{Planar two-dimensional analytical model}
%------------------------------------------------------------------------------
%
The mutual interaction of stellar winds is best and often modeled within the
framework of
either hydrodynamics (HD) or magnetohydrodynamics (MHD) (see, e.g., 
\citealt{Kissmann-etal-2016}). The purely hydrodynamical description of the 
subsonic interaction region allows  an analytical treatment based on the
theory of velocity or momentum density potentials. This concept has been 
introduced for the interaction of a stellar wind with the so-called
interstellar wind (blowing in the rest frame of a given wind-driving star as
a consequence  of the relative motion between star and surrounding interstellar
medium) by \citet{Parker-1961} \citep[see also][]{Nerney-Suess-1995}, and has
been used recently to compute the distortion of the local interstellar magnetic
field due to the solar wind plasma bubble
\citep{Roeken-etal-2015, Kleimann-etal-2016, Kleimann-etal-2017}.

In the following, we extend the modeling based on velocity potentials to the
case of multiple interacting stellar winds. In order to explore the principal 
structure of such multisource winds, we first analyze the planar 2D case
because it allows  the definition of a single stream function, and
thus an easy analytical representation of the stream lines of the
interacting winds. While a stream function is, in general, not available in
the 3D case \citep[see, e.g.,][]{Elshabka-Chung-1999, Lee-2009}, velocity
potentials can be defined without problem and  permit a semi-analytical
study, at least.      

%------------------------------------------------------------------------------
\subsection{Velocity potential and stream function of a multisource flow}
%------------------------------------------------------------------------------
%
The use of velocity potentials is based on the assumption
\begin{eqnarray}
  \label{eq:irrot}
  \nabla\times\vec{u} = \vec{0}
\end{eqnarray}
according to which flow velocity $\vec{u} = \vec{u}(\vec{r})$ at
\mbox{$\vec{r}=(x,y) \in \mathbb{R}^2$} is irrotational.
It has been demonstrated for the example of the interaction of the solar with
the interstellar wind that this is a very reasonable approximation for
steady-state flow configurations
\citep[e.g.,][and references therein]{Roeken-etal-2015}. For a generalization
to a momentum density potential see \citet{Kleimann-etal-2017}. The above
equation \eqref{eq:irrot} implies the existence of a velocity potential
$\Phi = \Phi(\vec{r})$ with
\begin{eqnarray}
  \vec{u} &=& \nabla\Phi
\end{eqnarray}
in which we adhere to the convention of omitting the minus sign. For the case
of a single point source of strength $m$ located at $\vec{r}_{\star}$, i.e.,\ an
isolated, noninteracting stellar wind, the velocity potential and the
associated stream function $\Psi = \Psi(\vec{r})$ are given by 
\begin{eqnarray}
  \Phi(\vec{r}) &=& m \ln\vert\vec{r}-\vec{r}_{\star}\vert 
  = \frac{m}{2} \ln \left[(x-x_\star)^2+(y-y_\star)^2\right] \\
  \label{eq:Psi_LL}
  \Psi(\vec{r}) &=& m \arctan \left(\frac{y-y_\star}{x-x_\star}\right) 
\end{eqnarray}
\citep{Landau-Lifschitz-1966}.
The neglect of the third Cartesian coordinate indicates that we limit our
consideration in this section to the 2D planar case. The flow velocity can 
be computed from either $\Phi$ or $\Psi$:
\begin{eqnarray}
  u_x &=& \frac{\partial \Phi}{\partial x} = \frac{\partial \Psi}{\partial y} 
  \label{cauchyriem1} \\
  u_y &=& \frac{\partial \Phi}{\partial y} =-\frac{\partial \Psi}{\partial x} 
  \label{cauchyriem2}
\end{eqnarray}
This dependence on the stream function implies
\begin{eqnarray}
  \label{eq:incompr}
  \nabla\cdot\vec{u} &=& \partial_x (\partial_y \Psi)
  + \partial_y (-\partial_x \Psi) = 0 
,\end{eqnarray}
and thus that the flow is incompressible and obeys the homogeneous mass
continuity equation. The significance of the functions $\Psi$ and $\Phi$ lies
in the fact that lines of constant $\Psi$ represent stream lines to which lines
of constant $\Phi$ are perpendicular, such that these lines span the grid of a
curvilinear, orthogonal, flow-aligned coordinate system.

Due to the linearity of Eqs.~\eqref{cauchyriem1} and \eqref{cauchyriem2},
the flow from $N$ different (point) sources of strength $m_k$ located at
$\vec{r}_k=(x_k,y_k)$ is obtained from a superposition 
\begin{eqnarray}
  \label{multiphi} \Phi(\vec{r}) &=&
  \sum\limits_{k=1}^N \frac{m_k}{2} \ln \left[(x-x_k)^2+(y-y_k)^2\right] \\
  \label{multipsi}
  \Psi(\vec{r}) &=&
  \sum\limits_{k=1}^N m_k \arctan \left(\frac{y-y_k}{x-x_k}\right) 
\end{eqnarray}
of the respective individual potentials and stream functions as
\begin{eqnarray}
  \label{multiux}
  u_x &=& \sum\limits_{k=1}^N \frac{m_k \, (x-x_k)}{(x-x_k)^2+(y-y_k)^2} \\
  \label{multiuy}
  u_y &=& \sum\limits_{k=1}^N \frac{m_k \, (y-y_k)}{(x-x_k)^2+(y-y_k)^2} \ .
\end{eqnarray}

\subsection{Complex potential} 
Upon noting from Eqs.~\eqref{cauchyriem1} and \eqref{cauchyriem2} that $\Phi$
and $\Psi$ fulfill the Cauchy--Riemann differential equations, one can define 
a complex potential at $z = x+\ii y \in \mathbb{C}$ via
\begin{eqnarray}
  w(z) &=& w(x+\ii y) = \Phi(x,y) + \ii \, \Psi(x,y) 
,\end{eqnarray}
and can thus combine Eqs.~\eqref{multiphi} and \eqref{multipsi} into one, 
\begin{eqnarray}
  \label{eq:w_complex1}
  w(z) &=& \sum\limits_{k=1}^N m_k \left[\ln\vert z-z_k\vert 
    + \ii \arctan \left(\frac{\Im(z-z_k)}{\Re(z-z_k)}\right)\right]
,\end{eqnarray}
with $\Re(\cdot)$ and $\Im(\cdot)$ denoting the real and imaginary part of
their complex argument. The analogous definition of a complex velocity
\begin{eqnarray}
  v(z) &=&
  \frac{\dd w}{\dd z} = \frac{\dd \Phi}{\dd x}+\ii \, \frac{\dd\Psi}{\dd x}
  = u_x - \ii u_y
  \label{complexvz1}
\end{eqnarray}
(note the minus sign!) enables us to further combine Eqs.~\eqref{multiux} and
\eqref{multiuy} into
\begin{eqnarray}
  v(z) &=& \sum\limits_{k=1}^N \frac{m_k\overline{(z-z_k)}}{\vert z-z_k\vert^2}
  = \sum\limits_{k=1}^N \frac{m_k}{z-z_k}
  \label{complexvz2}
\end{eqnarray}
with a bar indicating the complex conjugate.

A crucial consequence of choosing $\Psi$ according to Eq.~\eqref{eq:Psi_LL}
is that this choice implies
\begin{equation}
  \Psi(\vec{r}_\star-\vec{d})=\Psi(\vec{r}_\star+\vec{d})
\end{equation}
for any $\vec{d} \in \mathbb{R}^2$, and thus prevents us from distinguishing
pairs of stream lines on  opposite sides  sides of the source, since these get mapped
to the same $\Psi$ value. A viable method for circumventing this shortcoming is to
replace the arctan in Eq.~\eqref{eq:Psi_LL} by its two-argument form, giving
\begin{eqnarray}
  \label{eq:Psi_2pi}
  \Psi(\vec{r}) &=& m \arctan( y-y_\star, x-x_\star), \\
  &=& m \, \mathrm{arg}\left[ (x-x_\star) + \ii (y-y_\star) \right], \nonumber
\end{eqnarray}
where $\varphi=\mathrm{arg}(x + \ii y) \in [-\pi,\pi]$ is the unique angle
satisfying
\begin{equation}
  \label{eq:polar_coords}
  x = r \cos \varphi \quad \mbox{and} \quad y = r \sin \varphi
\end{equation}
for any $(x,y)$ with $r=\sqrt{x^2+y^2} > 0$.
In particular, we see that for the special case $\vec{r}_\star=\vec{0}$, the
stream function $\Phi$ is simply given by the polar azimuthal coordinate
$\varphi$, whose range is now $[-\pi,\pi]$ instead of the more conventional
$[0,2\pi]$. Furthermore, since we then recover the standard definition
\begin{eqnarray}
  \ln (z) &=& \ln |z| + \ii \, \mathrm{arg}(z)
\end{eqnarray}
of the complex-valued logarithm, the total complex potential
\eqref{eq:w_complex1} simplifies considerably to
\begin{eqnarray}
  w(z) &=& \sum\limits_{k=1}^N m_k \ln (z-z_k) \ ,
\end{eqnarray}
in full consistency with Eq.~\eqref{complexvz2}.

The use of the above potential theory for an exploration of the principal 
structure of the flow resulting from the interaction of multiple stellar winds 
is justified by the notion that this interaction occurs between the subsonic 
regions of the winds, which  in a very good approximation can be considered as
incompressible and irrotational \citep[e.g.,][]{Siewert-etal-2014, 
  Roeken-etal-2015}.
But even at larger distances from the cluster, where we expect the flow
velocity to approach a constant value, and thus mass density $\rho$ to
decrease with distance $r$ as $\rho(r) \propto r^{-2}$, the derived flow
structure is still valid. This can formally be seen by exchanging
\mbox{$\vec{u} \rightarrow \vec{U}:=\rho \vec{u}$}
\citep[see also][]{Kleimann-etal-2017} and noting that $\vec{U}$ and $\vec{u}$
share the same structure of stream lines. The possibly questionable
constraint of incompressibility \eqref{eq:incompr} now becomes
\mbox{$\nabla \cdot (\rho \vec{u})=0$} and merely implies mass conservation,
which is certainly not under dispute, but in fact constitutes a very desirable
property of any physical flow model.

%------------------------------------------------------------------------------
\subsection{Stream lines and boundary surfaces}
%------------------------------------------------------------------------------
\subsubsection{Discontinuity of the stream function}
Another technical problem in using the stream function as defined in
Eq.~\eqref{multipsi} is its discontinuity, which becomes evident when
considering the limits
\begin{eqnarray}
  \lim\limits_{x \searrow x_k} \Psi(x,y) &=&
  \Psi_{[k]}(x,y) \pm \frac{\pi}{2} \, m_k \label{limpsi1} \\
  \lim\limits_{x \nearrow x_k} \Psi(x,y) &=&
  \Psi_{[k]}(x,y) \mp \frac{\pi}{2} \, m_k \label{limpsi2}
\end{eqnarray}
with $\Psi_{[k]}(x,y)$ denoting the total stream function due to all sources
except $m_k$, assuming that none of them is located at $x=x_k$ as well,
i.e.,\ $\Psi$ is discontinuous in $(x_k,y)$ where it jumps by $\pm m_k \pi$.
For the calculation of stream lines via the equation $\Psi =$~const.,\ it is
therefore necessary to add a suitable term $\pm m_k \pi$ to the constant,
possibly multiple times (see below).
This inconvenience can be moderated (though not avoided entirely for principal
reasons) by using Eq.~\eqref{eq:Psi_2pi} in favor of Eq.~\eqref{eq:Psi_LL}.
Since the double-argument arctan covers the full range $[-\pi,\pi]$ rather
than just $[-\pi/2,\pi/2]$ as its single-argument form does,
Eqs.~\eqref{limpsi1} and \eqref{limpsi2} are then replaced by
\begin{eqnarray}
  \label{eq:single_jump}
  \lim\limits_{\substack{y \rightarrow y_k \\ y \gtrless y_k}} \Psi(x,y) &=&
  \Psi_{[k]}(x,y) \mp \pi \, m_k
\end{eqnarray}
such that there is just one single discontinuity to the left of $\vec{r}_k$ in
the $-x$ direction, across which $\Psi$ jumps by a full $\pm 2\pi \, m_k$.

For situations where the stream function is to be evaluated at some
position $y=y_\star$ that has multiple sources located to its right (also at
$y_k=y_\star$), Eq.~\eqref{eq:single_jump} may be generalized to
\begin{eqnarray}
  \label{eq:multi_jump}
  \lim\limits_{\substack{y \rightarrow y_\star \\ y \gtrless y_\star}} \Psi(x,y) =
  \sum_{k=1, k \not \in {\cal J}}^N m_k \, {\rm arctan}(y-y_k, x-x_k)
  \mp \pi \sum_{k \in {\cal J}} m_k &&
,\end{eqnarray}
where the position-dependent index set
\begin{eqnarray}
  {\cal J}(x,y) &=& \{ k \, | \, (x_k > x) \wedge (y_k = y) \}
\end{eqnarray}
serves to enumerate all sources to the right of the position in question
(and sharing the same $y$ coordinate).
\begin{figure*}[t!]
  \centering
  \includegraphics[width=0.9\textwidth]{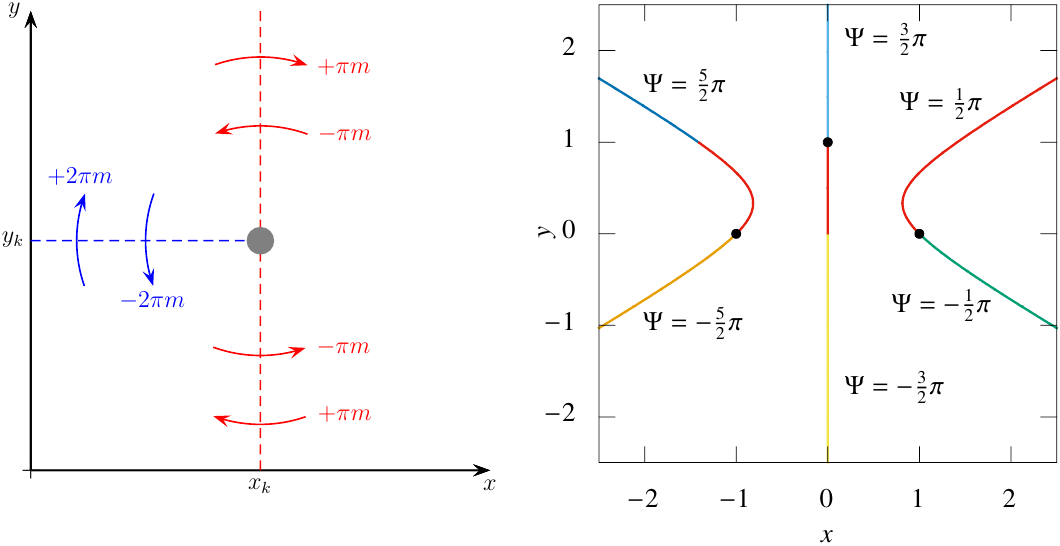}
  \caption{ \label{fig-jump}
    Left: Principal jump corrections of the stream function $\Psi$ induced by
    the presence of a source of strength $m$ (gray circle) located
    at $\vec{r}_k$.
    The use of the single-argument arctan \eqref{eq:Psi_LL} for the definition
    of $\Psi$ entrails jumps by $\pm \pi m$ (shown in red) when crossing the
    vertical line $x=x_k$ either ``above'' ($y>y_k$) or ``below'' ($y<y_k$)
    the source, while in the case of definition \eqref{eq:Psi_2pi}, only a
    single jump by $\pm 2\pi m$ (shown in blue) occurs across the horizontal
    line $y=y_k$, $x<x_k$ to the ``left'' of the source.
    Right: Sample stream lines for three stellar wind sources with equal
    strengths $m_1 = m_2 = m_3 = 1$ (black circle) computed from curves of
    constant stream function $\Psi(x,y)$ as given in Eq.~\eqref{eq:Psi_2pi}
    with jump corrections according to Eq.~\eqref{eq:multi_jump}. Colors
    indicate the various values of $\Psi$. The two stream lines emanating from
    the right source at $\vec{r}_3=(1,0)$ are both singled valued because there
    is no other source to its right, while the stream line pointing into the
    $-y$ direction from the central source at $\vec{r}_2=(0,1)$ has a jump of
    $-\pi$ at $y=y_3=0$. Finally, the source at $\vec{r}_1=(-1,0)$ has jumps of
    \mbox{$\pm \pi$} in both the upward- and downward-pointing stream line at
    $y=y_2=1$ and $y=y_3=0$, respectively.
  }
\end{figure*}
The need to carefully take these discontinuities into account is illustrated in
Fig.~\ref{fig-jump}, in the right panel showing sample stream lines for three
equally strong stellar wind sources. The stream lines are computed from
$\Psi(x,y)$ using the stream function \eqref{eq:Psi_2pi} with jump corrections
and labeled by their respective values of $\Psi$.
These corrections are necessary across the negative horizontal lines $y = y_k$
when using Eq.~\eqref{eq:Psi_2pi} (or, alternatively, the vertical line
$x = x_k$ when using Eq.~\eqref{eq:Psi_LL}) through the sources as illustrated
in the left diagram of the figure.

%------------------------------------------------------------------------------
\subsubsection{Boundary surfaces}
Assuming laminar flow conditions so that the plasmas of different stellar winds
cannot mix,  separate regions form in a steady state; these regions contain the outflowing
material from the different sources. Consequently, there are boundary
surfaces (or boundary lines in a plane) that separate the different stellar
wind flows. Those points on the boundaries where the flow
velocity vanishes and that are stream line-connected
with more than one source are called stagnation points. According to Eqs.~\eqref{cauchyriem1} and 
\eqref{cauchyriem2} these are the critical points of the stream function,
i.e.,\ those with 
\mbox{$\partial \Psi/\partial x = \partial \Psi/\partial y = 0$}.
With Eqs.~\eqref{complexvz1} and \eqref{complexvz2} this condition can be
written as 
\begin{eqnarray}
  \sum\limits_{k=1}^N \frac{m_k}{z-z_k} &=& 0 \ ,
\end{eqnarray}
which translates into 
\begin{eqnarray}
  \label{eq:sum_of_prod} \nonumber
  & \displaystyle\sum\limits_{k=1}^N m_k \ \frac{ 
    \prod\limits_{i=1,i \neq k}^N (z-z_i)}{\prod\limits_{i=1}^N(z-z_i)} & = 0 \\
  \Leftrightarrow
  & \displaystyle\sum\limits_{k=1}^N m_k \prod_{i=1,i \neq k}^N (z-z_i) & = 0
\end{eqnarray}
after some algebraic manipulations.

\begin{figure}[h!]
  \begin{center}
    \includegraphics{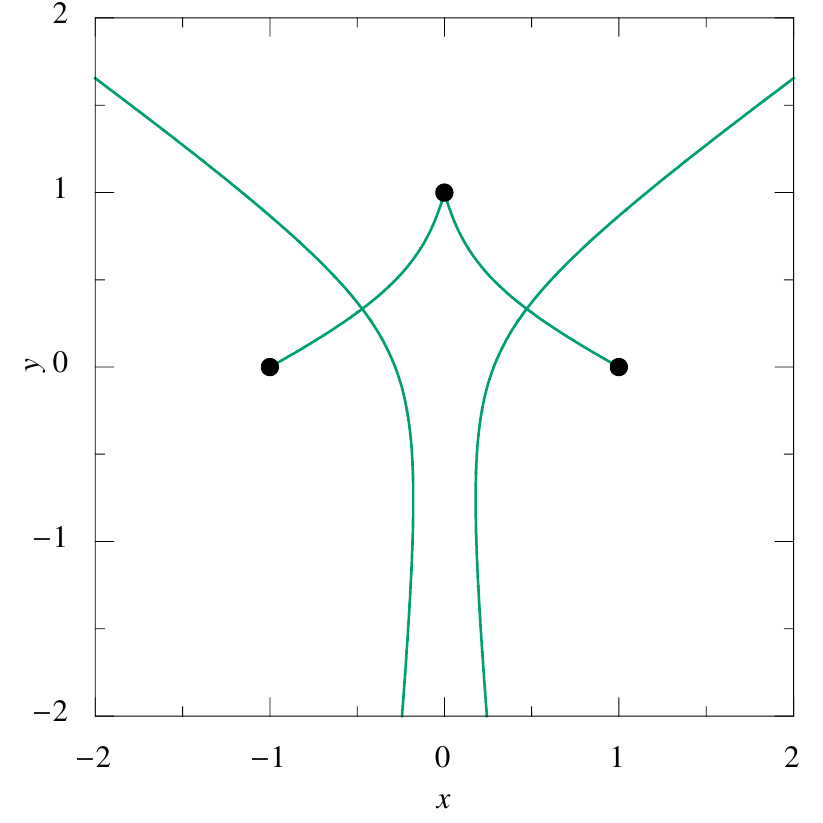}
  \end{center}
  ~\vspace*{-1.0cm}\\
  \caption{Stream lines through the two stagnation points of the
    configuration shown in Fig.~\ref{fig-jump}. The stagnation points are
    located where the stream lines seem to cross, although they actually do not.
  }
  \label{fig-stagnation}
\end{figure}

The three-star configuration shown in Fig.~\ref{fig-jump} exhibits two
stagnation points, namely where the associated stream lines seem to cross
(see Fig.~\ref{fig-stagnation}). Of course, they do not actually cross:
the flows from the stars decelerate to zero velocity towards a stagnation
point, so that a line connecting two sources represents two stream lines with
opposite flow directions corresponding to the outflow from each star. The other
two lines, not connected to the stars, indicate whereto the material is flowing
away from the stagnation points: these lines represent the boundaries
separating the nonmixing stellar wind plasmas. 

From this symmetric configuration of sources in the corners of an isosceles
triangle we can already draw some conclusions. First, it becomes clear that
the resulting large-scale wind has individual regions where the outflow remains
determined by the star whose material is filling it. Second, it is likely that
comparatively narrow {outflow channels} are forming, 
one of which is visible in Fig.~\ref{fig-stagnation}; some of the wind material
of the ``upper'' star escapes from the cluster  into the region
towards positive $y$-values, and also within a narrow ``channel'' forming
between the wind regions of the other two stars. 

%------------------------------------------------------------------------------
\subsubsection{Asymptotic flow behavior}
\label{sec:asymptotics}
\begin{figure}[h!]
  \begin{center}
    \includegraphics{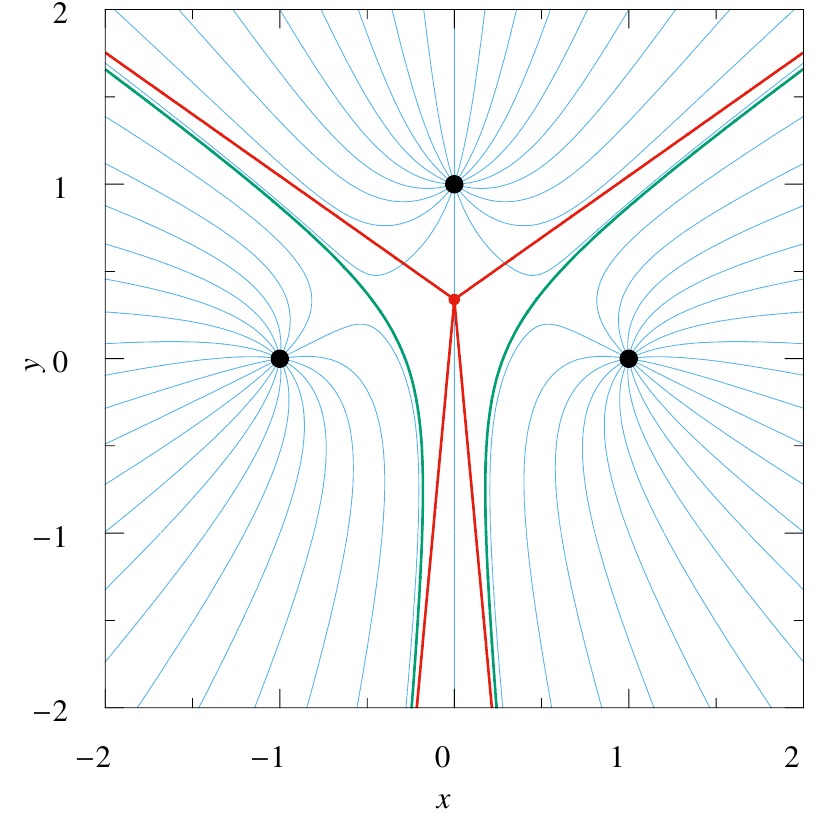}
  \end{center}
  ~\vspace*{-1.0cm}\\
  \caption{Resulting flow of the configuration of stellar wind sources
    shown in Fig.~\ref{fig-jump} visualized by many stream lines (blue).
    In addition, the two boundary lines (green) are indicated, as are  their
    asymptotic directions (red). The corresponding straight lines cross in the
    origin (Eq.~\eqref{origin}) of the shifted polar coordinate system.}
  \label{fig-isosceleflow}
\end{figure}

The above findings motivate an analysis of the asymptotic flow behavior. To this end, it is advantageous to transform from Cartesian to polar coordinates
\eqref{eq:polar_coords}, such that 
\begin{eqnarray} 
  \label{eq:Psi_asympt}
  \nonumber  \Psi(r,\varphi) &=& \sum_{k=1}^N m_k
  \arctan \left(r \sin\varphi - y_k, r \cos\varphi - x_k\right) \\ 
  &=& \sum_{k=1}^N m_k \arctan
  \left(\sin\varphi - y_k/r, \cos\varphi - x_k/r\right)  \\
  \nonumber  & & \xrightarrow{r \gg \vert{x_k}\vert,\vert{y_k}\vert }
  \left(\sum_{k=1}^N m_k\right) \varphi
\end{eqnarray}
We note that for the last step, the use of the double-argument arctan
\eqref{eq:Psi_2pi} is mandatory;  otherwise,
\mbox{$\arctan[\tan(\varphi)]=\varphi$} would only hold for the left $(x\ge 0)$
half-plane, in which $\varphi \in [-\pi/2,\pi/2]$. Thus, for a given
jump-corrected value $c$ of a boundary stream line, we obtain from the stream
line equation $\Psi = c$ the angle of its asymptotic direction:
\begin{eqnarray}
  \varphi &=& c\,\left(\sum\limits_{k=1}^N m_k\right)^{-1} 
\end{eqnarray}
Choosing a new origin of the polar coordinate system in the location of the
cluster's barycenter
\begin{eqnarray}
  \vec{r}_0 &=& \left( \sum\limits_{k=1}^N m_k \vec{r}_k \right) \,
  \left( \sum\limits_{k=1}^N m_k \right)^{-1}
  \label{origin}
\end{eqnarray}
gives the asymptotic lines for the overall flow configuration shown in
Fig.~\ref{fig-isosceleflow}. 

%------------------------------------------------------------------------------
\subsection{Principal structure of the large-scale cluster wind}
\begin{figure}[h!]
  \begin{center}
    \includegraphics{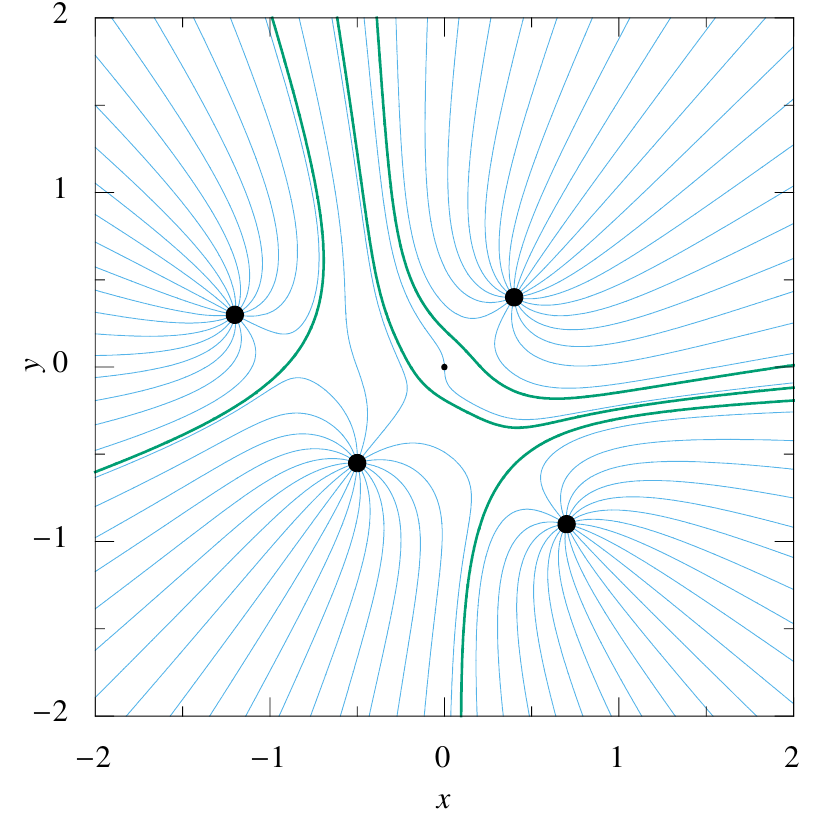}
  \end{center}
  ~\vspace*{-1.0cm}\\
  \caption{Flow configuration resulting from five stellar wind sources,
    with the one star in the origin having 1/10 of the strength of the other
    four.} 
  \label{fig-fivestars}
\end{figure}

%------------------------------------------------------------------------------
%
To conclude the treatment of planar configurations, we consider a more
realistic, asymmetric configuration of five stars with one wind having only
$1/10$ of the source strength of the other four (see Fig.~\ref{fig-fivestars}).
While the actual flow field is more complicated, its principal structure is
analogous to that discussed for the symmetric configuration above. There are
five outflow regions, each associated with one of the five sources.
The total angular extent covered by any of these regions can either form a
continuous interval or be partitioned into at most $N-1$ smaller
subintervals, some of which may take the form of narrow outflow channels.
In the example shown in Fig.~\ref{fig-fivestars}, the source in the bottom left
quadrant has two outflow channels and one large outflow region, while the
small central source also has two channels but no large outflow region, and
the other three sources each have a continuous outflow region and no channels
at all.
A simple example featuring the maximum number of outflow channels would be a
set of $N-1$ sources of uniform strength placed equidistantly on a circle,
together with an additional source of arbitrary strength located at the
circle's center. Specifically, if the four equally strong sources depicted in
Fig.~\ref{fig-fivestars} were located in the corners of a square $[-1,+1]^2$,
the setting's symmetry would force the four outflow channels to be located
around the Cartesian $x$- and $y$-axes, carrying material from the central
source into the $\pm x$ and $\pm y$ directions.)

It is interesting to ask what the velocity field far from a cluster looks like.
As in the case for the asymptotic directions discussed in the previous section,
this information can be obtained from the asymptotics of the combined velocity
potential \eqref{multiphi}. Using again polar coordinates, we obtain
\begin{eqnarray}
  \nonumber \Phi(r,\varphi) &=& \sum_{k=1}^{n} \frac{m_k}{2}
  \ln \left[(r \cos\varphi -x_k)^2+(r \sin\varphi -y_k)^2\right] \\
  &=& \sum_{k=1}^{n} \frac{m_k}{2} \left(2\ln r +
  \ln \left[\left(\cos\varphi - \frac{x_k}{r}\right)^2+\left(\sin\varphi -\frac{y_k}{r}\right)^2\right]\right) \nonumber \\
  & & \xrightarrow{r \gg \vert {x_k}\vert ,\vert {y_k}\vert }
  \left(\sum_{k=1}^{n} m_k\right) \ln r
\end{eqnarray}
Evidently, the outflow regions are adjusting to each other in such a way that
the asymptotic large-scale wind is given by that of a point source with
strength $\sum_{k=1}^{n} m_k$, i.e.,\ the combined source strengths of all
wind-driving stars in the cluster. Consequently, at large distances the
cluster outflow is approximately spherically symmetric. 

%------------------------------------------------------------------------------
\section{Extension to three dimensions}
%------------------------------------------------------------------------------
%
Despite the valuable insights obtained from the analytical modeling discussed
so far, the major drawback of this model lies in its limitation to a planar 2D
configuration.
While it could be argued that the topological structure of a 2D flow model will
be very similar, if not identical, to that of a 3D model where both the
distribution of sources and the region of analysis is limited to a plane
(provided that only locations within this plane are considered), there will of
course be quantitative differences. Most notably, the flow speed $|\vec{u}|$ of
a 2D point source (which actually corresponds to a line source in 3D) decreases
with radius as $r^{-1}$, rather than $r^{-2}$. For these reasons, we now turn to
the case of asymmetric arrangements of point sources in full 3D.

While it is of course still possible to compute the overall flow field from the
superposition of the individual velocity potentials of the sources distributed
in 3D, the easy access to stream lines via a stream function is no longer
available in that case, and a numerical procedure becomes necessary.
Rather than improving the model further by actually computing the
interaction of multiple stellar winds on the basis of numerical solutions of
the (M)HD equations (which we intend to study in future work), we now describe
this procedure, the underlying mathematical theory, and the results thus
obtained.  
 
% ------------------------------------------------------------------------------
\subsection{Mathematical background}
\label{sec:maths}
%------------------------------------------------------------------------------
%
The statements from the previous section generalize to 3D as follows.
For a set of $N$ point sources with strengths $m_k$ and locations
\mbox{$\vec{r}_k = (x_k, y_k, z_k)$} in 3D space, the total velocity field
found from the superposition
\begin{eqnarray}
  \label{eq:total_v_from_pot}
  \vec{u}(\vec{r}) &=& \nabla \left(
  \sum\limits_{k=1}^N \frac{m_k}{|\vec{r}-\vec{r_k}|} \right)
\end{eqnarray}
can have at most $N-1$ stagnation points (also called  nulls). In the vicinity of a
null at $\vec{r}_{\rm null}$, $\vec{u}(\vec{r})$ can be approximated by
\begin{eqnarray}
  \vec{u}(\vec{r}) &=& \underbrace{\vec{u}(\vec{r}_{\rm null})}_{= \, \vec{0}}
  + \, \J(\vec{u})|_{\vec{r}_{\rm null}} \, (\vec{r}-\vec{r}_{\rm null})
  + {\cal O} \left( |\vec{r}-\vec{r}_{\rm null}|^2 \right) 
,\end{eqnarray}
where $\J(\vec{u})|_{\vec{r}_{\rm null}}$ is the Jacobian matrix of $\vec{u}$ with
elements
\begin{eqnarray}
  [\J(\vec{u})]_{ij} &=& \frac{\partial u_i}{\partial r_j}
\end{eqnarray}
evaluated at the null.
Since $\vec{u}$ derives from a potential, we see from
\begin{eqnarray}
  [\J(\vec{u})]_{ij}
  &=& \frac{\partial}{\partial r_j} \frac{\partial \Phi}{\partial r_i}
  = \frac{\partial}{\partial r_i} \frac{\partial \Phi}{\partial r_j}
  = [\J(\vec{u})]_{ji}
\end{eqnarray}
that $\J$ is symmetric, implying that all of its eigenvalues are real, and that
the corresponding eigenvectors are orthogonal (or at least can be chosen to
have that property).
Furthermore, the trace
\begin{eqnarray}
  \mathrm{Tr} \, [\J(\vec{u})]
  &=& \sum\limits_{i=1}^d \frac{\partial u_i}{\partial r_i}
  = \nabla \cdot \vec{u}
\end{eqnarray}
of $\J$, i.e., the sum of its eigenvalues, vanishes for incompressible flow in
any dimension $d$. These constraints limit the topological variety that would
otherwise be possible for nulls in
the
following cases \citep[see, e.g.,][]{Parnell-etal-1996}:
\begin{itemize}
\item In 2D, only X-type nulls are possible, i.e., stream lines in the
  direction (anti)parallel to the negative eigenvalue's eigenvector converge
  towards the null (thus connecting it to either a source or another null),
  and an additional pair of stream lines emanate from the null into the
  perpendicular direction (as shown in Fig.~\ref{fig-stagnation}).
  The additional stream lines form separatrices, separating regions filled by
  material from different sources. The second possibility of an O-type null
  (with circular stream lines closing in on the null in their common center)
  is ruled out by $\nabla \times \vec{u}=\vec{0}$.
\item In 3D, the situation is similar. We also get individual ``spine'' stream
  lines that originate at a source and converge on the null (in the direction
  of the negative eigenvalue's eigenvector), but now the remaining two
  eigenvectors (whose eigenvalues are both positive) span an entire ``fan''
  plane that is orientated perpendicular to the spine. The set of stream lines
  starting from the null within this plane form separatrix surfaces, which
  again separate the regions that are filled by material from different
  sources.
\end{itemize}

It is also reasonable to generalize the reasoning presented in
Section~\ref{sec:asymptotics} for the 3D case. At large distances from the
cluster of point sources, the total velocity field approaches that of a
monopole (i.e., a single point source whose strength is the sum of the
individual strengths of the contributing sources), and when placing a
sufficiently large sphere around the cluster's barycenter, the wind from any
source $j$ fills a solid angle $\mu_j$ on that sphere given by
\begin{eqnarray}
  \label{eq:asmpt-3d}
  \frac{\mu_j}{4\pi} &=& m_j \, \left(\sum\limits_{k=1}^N m_k \right)^{-1}
\end{eqnarray}
irrespective of how complex the spatial arrangement of both the sources in
the cluster and their corresponding regions on the sphere may be.
This finding, which seems surprising at first sight,  is illustrated in the right panel
of Fig.~\ref{fig:maple3d+patches} and quantitatively supported by the data
listed in Table~\ref{tab:solid_angles}.

%------------------------------------------------------------------------------
\subsection{Finding null points and separatrices in three dimensions}
\begin{figure*}[t]
  \centering
  \begin{tabular}{cc}
    \includegraphics[width=0.45\textwidth]{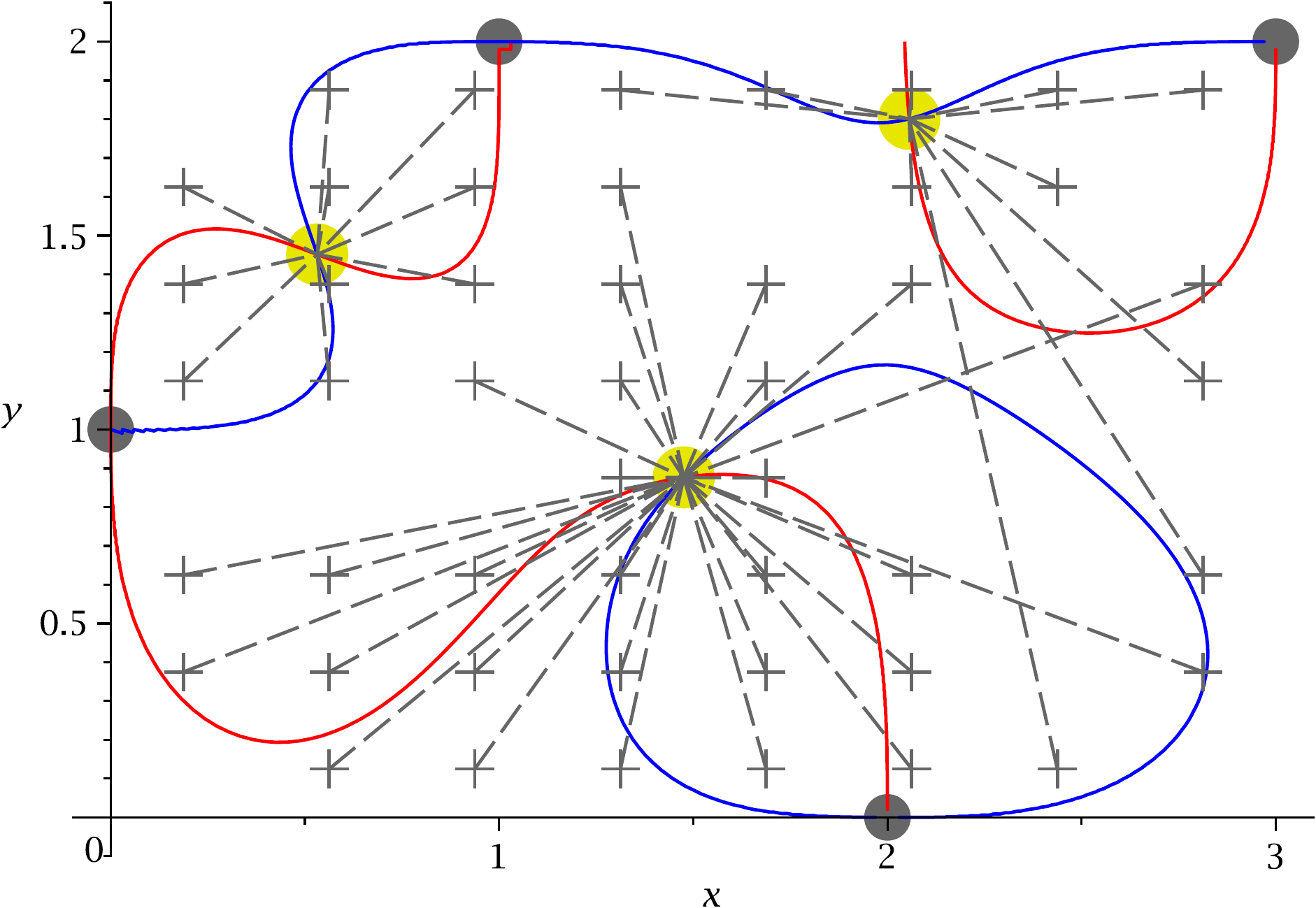} &
    \includegraphics[width=0.45\textwidth]{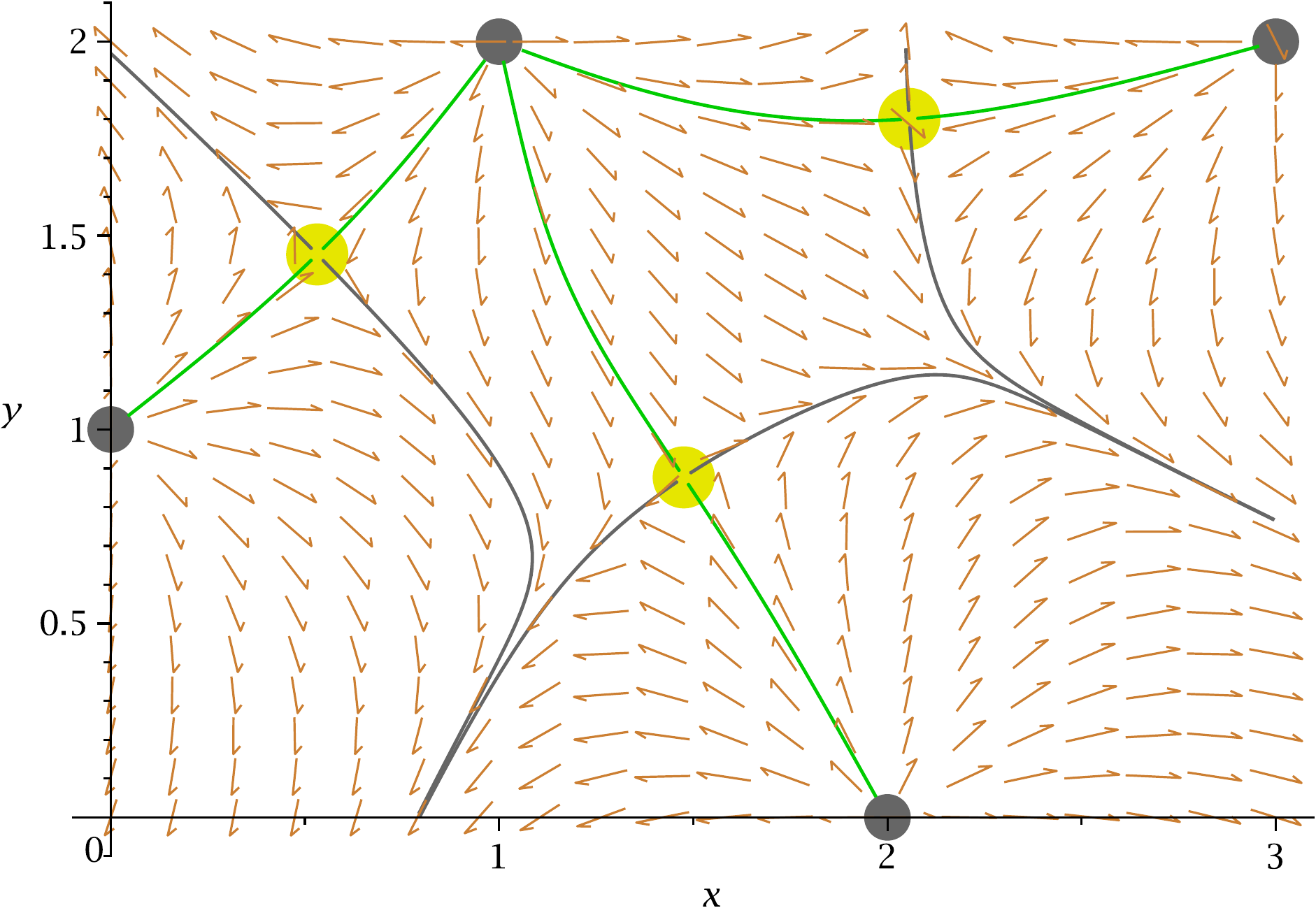}
  \end{tabular}
  \caption{Left: Curves on which the $x$ component (red) and the $y$ component
    (blue) of the combined velocity field of four sources (filled gray circles)
    vanish. Null points (filled yellow circles) are identified as those
    locations where both lines meet outside of any source. Straight dashed
    lines connect those initial centers $\vec{c}$ (crosses) of a rectangular cell that
    did eventually converge onto a null to this null (which apparently need
    not be the null closest to $\vec{c}$'s initial position). We also note  that
    these dashed lines merely serve to illustrate  which starting point
    converges upon which null, while convergence itself usually does not
    proceed along straight lines.
    Right: Same situation, indicating identified nulls
    at $(0.5316, 1.4514)$, $(1.4755, 0.8766)$, and $(2.0562, 1.8008$),
    separatrices (gray), spine stream lines (green), and unit vectors of the
    actual flow field $\vec{u}$ (brown arrows).
  }
  \label{fig:maple-2d}
\end{figure*}

%------------------------------------------------------------------------------
%
Since the structure of a given flow configuration tends to be more difficult to
visualize in 3D, it is important for the analysis to identify the ``skeleton''
consisting of critical field lines, which are those of the the spine and the
fan. The first step towards this goal is to locate all the stagnation points.
Since the structure of $\vec{u}$ resulting from
Eq.~\eqref{eq:total_v_from_pot} is not conducive to an analytical approach
(not even in 2D, where solving Eq.~\eqref{eq:sum_of_prod} amounts to finding
all complex roots of an $N$th-order polynomial), we resort to numerical
methods. The following procedure, which is a generalization of the
Newton--Raphson method, has shown to yield good results for the purpose at hand (for the sake of clarity, Fig.~\ref{fig:maple-2d} illustrates the situation in
a 2D plane, but the procedure in 3D is completely analogous):
\begin{enumerate}
\item A rectangular region containing all sources is partitioned into an
  \mbox{$(n \times n \times n)$} array of brick-shaped cells of equal size,
  where typically $n \lesssim 10$.
\item At the center $\vec{c}$ of each cell, each velocity component
  $u_{\alpha}$, \mbox{$\alpha \in \{x,y,z \}$} is approximated by its
  first-order Taylor polynomial
  \begin{eqnarray}
    p_{\alpha}(x,y,z) &=& u_{\alpha}|_{\vec{c}}
    + (x-c_x) \, (\partial_x u_{\alpha})|_{\vec{c}} \\ && \nonumber
    + (y-c_y) \, (\partial_y u_{\alpha})|_{\vec{c}}
    + (z-c_z) \, (\partial_z u_{\alpha})|_{\vec{c}}
  \end{eqnarray}
  and the position $\vec{c}_0$ is determined at which $p_{\alpha}(\vec{c}_0)=0$
  for all $\alpha$. If $\vec{c}_0$ is ill-defined, which can happen if the
  subspaces defined by the zeroes of any two polynomials are parallel, the cell
  is discarded, else it is centered on $\vec{c}_1$ and reduced in size by a
  factor of 1/2 in all directions.
\item Step 2 is repeated with $\vec{c}_0$ as the new $\vec{c}$ until the cell
  size has shrunk below a pre-defined threshold (typically $\lesssim 10^{-8}$).
\item The final $\vec{c}_0$ is written into a list of prospective nulls (if it
  does not coincide with a source), and the process is repeated for the next
  cell.
\item Finally, the list needs to be cleared of duplicates that occur
  whenever two cells converge upon the same null. The list must be
  checked that it contains exactly $N-1$ distinct points at which $\vec{u}$ is indeed
  sufficiently close to zero.
\end{enumerate}
This procedure was preferred over the topological degree method
\citep{Greene-1992} because it is  easier to implement. It is similar to the trilinear
method proposed by \citet{Haynes-Parnell-2007} except that the field components
within a cell are assumed to be planar, rather than trilinear, and that a given
null may  find multiple times from different starting
cells (and in fact usually does). We also skip the initial preconditioning of cells since for small cell
arrays the gain in execution speed does not justify the extra coding effort.

%------------------------------------------------------------------------------
\subsection{Numerical example}
%------------------------------------------------------------------------------
%
From the list of nulls thus identified, we can continue along the lines of
Section~\ref{sec:maths} by numerically evaluating the Jacobian and its
eigenvectors at each null, and then trace stream lines along the spine and into
selected directions within the fan plane.
As an instructive example, we consider the configuration specified in
Table~\ref{tab:solid_angles}, which consists of four unequally strong sources
forming an irregular tetrahedron, together with a fifth source placed in their
common barycenter at $\vec{r}_5=\vec{0}$.
As expected, four nulls are found, situated approximately between the
central source and each of the four vertices of the tetrahedron.

\begin{figure*}
  \centering
  \begin{tabular}{cc}
    \includegraphics[height=7cm]{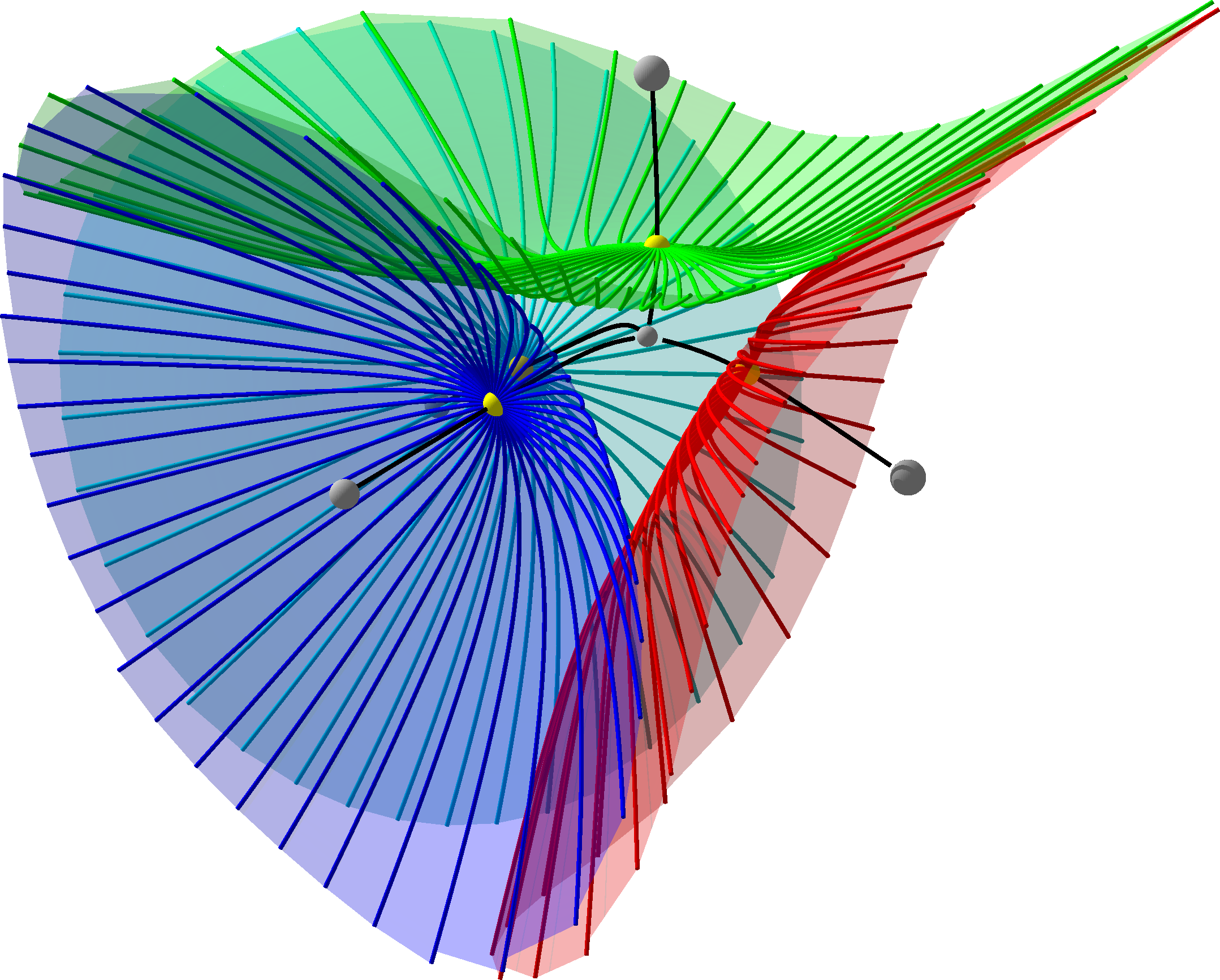} &
    \includegraphics[height=7cm]{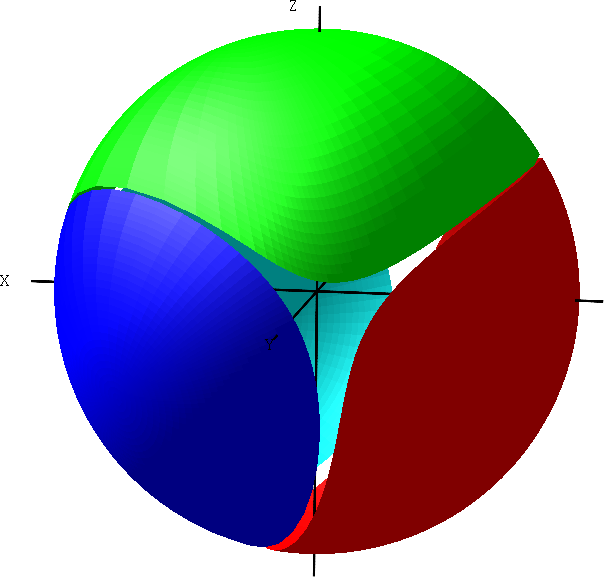}
  \end{tabular}
  \caption{ \label{fig:maple3d+patches}
    Left: 3D rendering showing the five sources (gray spheres, volumes
     proportional to  strength of the source),  four null points
    (yellow spheres),  spine stream lines (black lines), and  four
    separatrix surfaces (semi-transparent blue, red, green, cyan) emanating
    from the nulls, together with some of their constituting fan stream lines.
    Only parts of objects within a sphere of radius 5 around the origin are
    shown.
    Right: Perspective projection of a spherical surface of radius $R=10$,
    color-coded by the source from which stream lines passing through a given
    point on the sphere originate, thus indicating the extent and geometrical
    shape of the solid angle that is filled by the wind from the corresponding
    stellar source. The colors of sources 1 (cyan), 2 (blue), 3 (red), and 4
    (green) correspond to those of the null (as used in the left image) to
    which the respective source is connected by a streamline. The patch for
    the central source (filling the small lanes between the other four patches)
    is omitted.
  }
\end{figure*}

Having located the four stagnation points, we can generate a pair of stream
lines in the spine direction of each null, as well as a range of stream lines
(80 for each null in this case) starting in the fan plane by numerical
second-order Runge--Kutta integration. The resulting structural skeleton is
shown in the left plot of Fig.~\ref{fig:maple3d+patches}.
As can be seen, the separatrices form umbrella-like shapes around the four
outer sources, approaching the shape of wide cones at larger distances.

\begin{table}[h]
  \centering
  \begin{tabular}{c@{\hspace{3mm}}c@{\hspace{2mm}}c|ccc|c}
    \multicolumn{3}{c|}{Sources} & \multicolumn{3}{c|}{Fraction of $4\pi R^2$ filled at} & $\mu_j/(4\pi)$ \\
    $j$ & $m_j$ & $\vec{r}_j$ & $R=5$ & $R=10$ & $R=20$ & \\ \hline
    1 & 1.0 & $( 2,  -3,  -1)$ & 0.1077 & 0.1229 & 0.1247 & 0.1250 \\
    2 & 1.5 & $( 2,\, 2,  -1)$ & 0.1778 & 0.1858 & 0.1870 & 0.1875 \\
    3 & 2.5 & $(-2,\, 0,  -1)$ & 0.3141 & 0.3116 & 0.3114 & 0.3125 \\
    4 & 2.5 & $( 0,\, 0,\, 2)$ & 0.3230 & 0.3138 & 0.3121 & 0.3125 \\
    5 & 0.5 & $( 0,\, 0,\, 0)$ & 0.0774 & 0.0659 & 0.0648 & 0.0625 \\ %\hline
    $\sum_j$ & 8.0 &    ---    & 1.0000 & 1.0000 & 1.0000 & 1.0000 \\ \hline \\
  \end{tabular}
  \caption{ \label{tab:solid_angles}
    Strengths and positions of the five sources used in the example of
    Fig.~\ref{fig:maple3d+patches}, with the fraction of the total area of a
    sphere of radius $R$ filled by the stream lines emanating from that source
    vs.\ the solid angle divided by $4\pi$  expected for
    \mbox{$R \rightarrow \infty$} according to Eq.~\eqref{eq:asmpt-3d}.}
\end{table}

We use this example to confirm the prediction of Eq.~\eqref{eq:asmpt-3d}.
For this purpose, we construct the contour of points at which each of our fan
stream lines pierces a sphere of radius $R$ (see the right plot of
Fig.~\ref{fig:maple3d+patches}, which provides a visual impression of these
polygonal patches), and determine the area which these polygons encompass on
the sphere. As can be seen from the data presented in the right part of
Table~\ref{tab:solid_angles}, the solid angles thus computed are indeed
consistent with the predicted asymptotes to an excellent degree.

%------------------------------------------------------------------------------
\section{Summary and conclusions}
%------------------------------------------------------------------------------
%
With the present study we addressed the problem of the mutual interaction of
multiple winds from stars located within a stellar cluster. While several
studies have investigated large-scale cluster winds resulting from an
intra-cluster interaction of multiple stellar winds, they have as yet not
provided details of the colliding flows inside a given cluster. The present
work represents a further step towards a quantification of the principal
structure of the combined flow within and near such cluster.
To this end we have applied the theory of complex velocity potentials
to determine the stagnation points and the separatrices between the individual
stellar wind regions originating from a planar arrangement of point sources, as
well as the structure of the asymptotic large-scale wind outside the cluster.
While in two spatial dimensions the use of complex-valued potentials  allows 
a concise analytical treatment of stream lines using stream functions, the
extension to the more realistic 3D case requires recourse to numerical methods
(although the general theory of vector-valued nulls may be explored
analytically, as has in fact been done by various authors).

The main results can be summarized as follows. First, one can distinguish 
distinct regions in the large-scale cluster wind that are filled with plasma
from the individual stars. Second, as a consequence of the strengths and
locations of the wind sources, the outflow of a given star when projected onto
a sphere does not necessarily result in a singly connected region. The wind of
a given star may be forced, partly or fully, into comparatively narrow outflow
channels. Third, as expected, one can demonstrate analytically that the
large-scale cluster wind asymptotically approaches spherical symmetry at large
distances. 

In conclusion, we state that the combined flow inside a stellar cluster
resulting from the interaction of multiple stellar winds is highly structured.
This confirms the expectation, and is the motivation to carry out  further studies 
on the basis of (magneto)hydrodynamical simulations.
\begin{acknowledgements}
  We are grateful to the \textit{Deutsche Forschungsgemeinschaft (DFG)}
  for the funding project SCHE334/9-2.
  Furthermore, JK acknowledges financial support through the
  \textit{Ruhr Astroparticle and Plasma Physics (RAPP) Center}, funded as
  MERCUR project St-2014-040.
\end{acknowledgements}
% 
%\bibliographystyle{aa}
%\bibliography{references}
%

\end{document}